\documentclass[acmsmall]{acmart}
\usepackage{graphicx}
\usepackage{subcaption}
\AtBeginDocument{%
  }

\renewcommand\footnotetextcopyrightpermission[1]{}

\begin{document}

\title{How Students (Mis)understand Conditionals and Loops – A Taxonomy}

\author{Dimitri Eckert}
\email{dimitri.eckert@tuhh.de}
\orcid{0000-0003-0512-4611}
\affiliation{%
  \institution{Hamburg University of Technology}
  \city{Hamburg}
  \country{Germany}
}

\author{Christian Kautz}
\email{kautz@tuhh.de}
\orcid{0000-0001-9665-8162}
\affiliation{%
  \institution{Hamburg University of Technology}
  \city{Hamburg}
  \country{Germany}
}

\renewcommand{\shortauthors}{Eckert et al.}

\begin{abstract}
Understanding student difficulties in programming is a complex challenge due to the wide range of topics and the abundant varieties of misconceptions and errors. This paper presents the design and development of a fine-grained taxonomy that categorizes novice programmers’ difficulties specifically related to reading and understanding the control flow constructs selection and iteration. Building upon prior research and our own empirical data from quizzes and interviews with students, the taxonomy is constructed through the iterative methodology of the Extended Taxonomy Design Process (ETDP). Key contributions include clear distinctions between different student difficulties and a detailed analysis of common student misunderstandings concerning conditional statements and loops. The taxonomy aims to aid computing education researchers by providing a harmonized framework to classify and analyze student errors, fostering deeper theoretical insights and informing pedagogical strategies. Future work will involve applying the taxonomy to novel student data and evaluating its usability among educators and researchers.
\end{abstract}





\maketitle

\section{What is a taxonomy?}
According to \cite{kundisch_update_2022}, taxonomies are "classification systems that help researchers conceptualize phenomena based on the dimensions and characteristics of these systems". Fundamentally, humans categorize to manage the overwhelming diversity of objects they encounter, facilitating sensemaking and theory building \cite{nickerson_method_2013}. Taxonomies serve as foundational artefacts in this context, acting as structured models that represent domains through relevant constructs and their interrelationships \cite{march_design_1995}. These artifacts adhere to essential model properties: representation (depicting existing or potential objects), reduction (focusing on attributes pertinent to a taxonomy’s purpose and audience), and pragmatism (serving various functions such as description or analysis) \cite{stachowiak_allgemeine_1973}. Understanding students conceptions about programming concepts is a difficult task. Even more so as on one hand the range of topics covered in typical programming courses is very broad and on the other also the difficult aspects per topic are abundant. A taxonomy can therefore help to get a better overview, add structure to the discussions about student difficulties and highlight relevant clusters and dimensions of difficulties.

\section{Definition of misconceptions, difficulties and errors}
The terms "misconception", "difficulty" and "error" have different meanings in various contexts and are sometimes used interchangeably. To avoid confusion, we describe here what we mean by these terms. We use the term "error" to describe a concrete incidence of a student making a mistake while trying to understand a program, leading to wrong assumptions about how that program behaves. We use the term "difficulty" for a misunderstanding, a lack of understanding or an erroneous cognitive behavior with regard to a concept or aspect that causes students to make errors. Errors are therefore the materialization of difficulties. Lastly, a "misconception" is a specific kind of difficulty that is based in a concrete mental model about a specific programming concept that does not align with reality or widely accepted scientific notions and therefore leads to errors. \cite{qian_students_2017, cortinovis_open_2023, swidan_programming_2018, shah_qualitative_2017} We also follow the description of misconception by \cite{swidan_programming_2018}, as being non-binary. Students can hold semi-correct or multiple (correct or incorrect) notions of a concept at the same time. In order to have a misconception students need to have some form of understanding of a concept even though it is incorrect or deficient. On the other hand, we challenge the definition by \cite{cortinovis_open_2023} which states that misconceptions are necessarily stable and systematically repeated. Especially in programming where novice students usually have little to no prior knowledge misconceptions can be fragile knowledge held by the students trying to make some sense of a program, as opposed to firm believes about a concept (as for example in physics education).

\section{Scope} \label{scope}
In this paper we focus on the difficulties concerning the process of reading and understanding code. Therefore, we do not consider any difficulties that solely arise while writing code. Furthermore, we only cover difficulties concerning conceptual knowledge (as opposed to syntactic and strategic knowledge \cite{bayman_using_1988, mcgill_conceptual_1997, qian_students_2017}) connected to the concept of control flow of programs containing the control structures iteration and/or selection. The literature we considered, as well as our own research, focuses on imperative programming languages such as C, C++, Python and Java, as well as block-based languages like Scratch. We only consider errors stemming from outright misconceptions or difficulties understanding the programming constructs. This means that we do not include syntactic or lexical errors as we consider them to be fundamentally different in nature.

\section{Methods} 
We follow the method of the extended taxonomy design process (ETDP) as described by \cite{kundisch_update_2022}. This method is an extension to the method developed by\cite{nickerson_method_2013}. 

\begin{figure}
    \centerline{\includegraphics[width=5.1in]{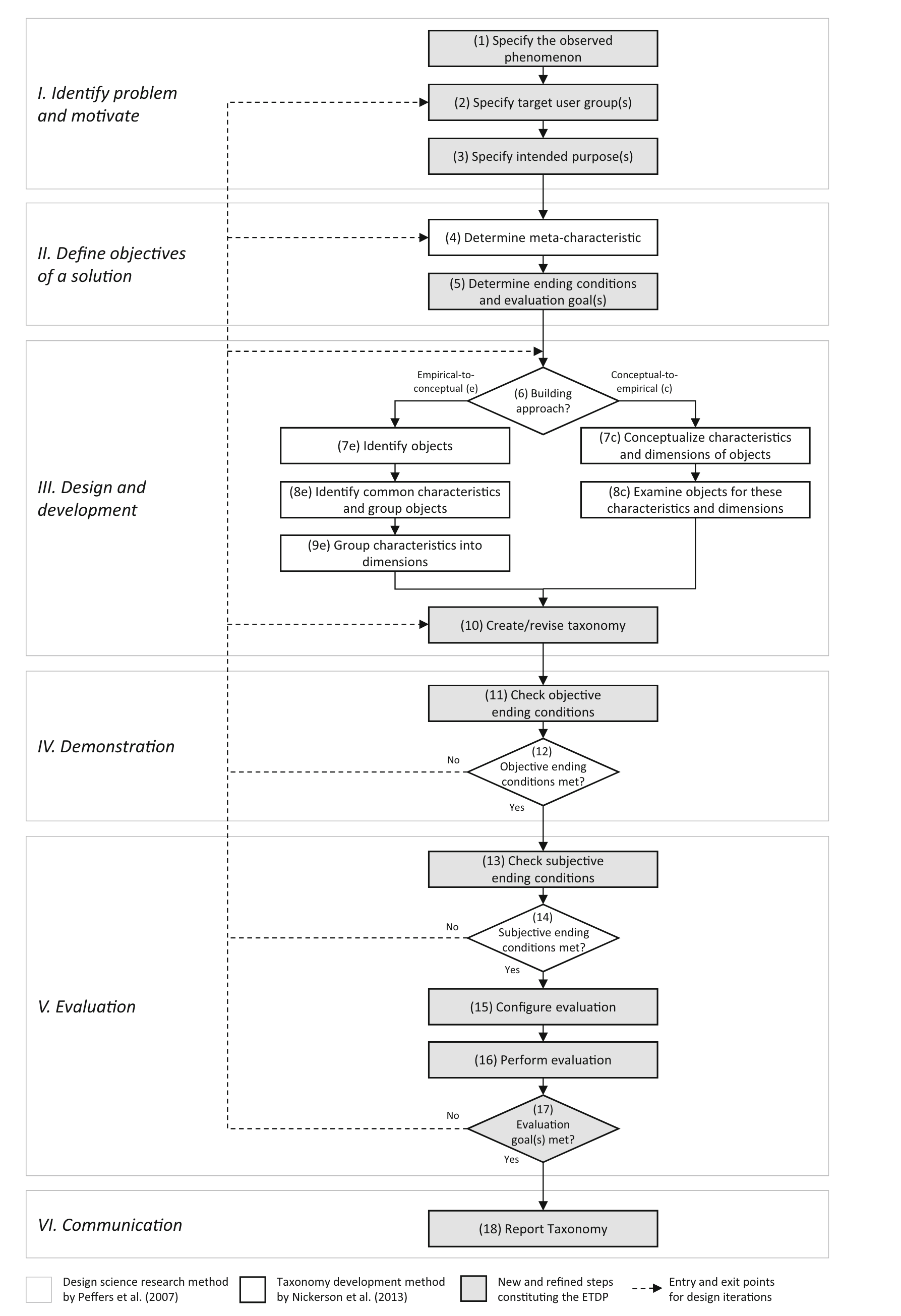}}
    \caption{The "Extended taxonomy design process (ETDP)" as described by \cite{kundisch_update_2022}}
    \label{fig:KundischProcess}   
\end{figure}

The ETDP is inherently iterative, allowing taxonomy designers to continuously update, extend, and refine existing taxonomies without restarting the entire design procedure. This flexibility is crucial for adapting taxonomies to evolving research contexts, which are shaped by the specific phenomenon under consideration, the target user groups, and the intended purposes of the taxonomy. This also applies to the domain of programming difficulties, as the requirements for a taxonomy can change depending on, for example, the age and prior knowledge of the students under consideration. Additionally, the iterative nature of the process allows for relatively easy incorporation of future research that is bound to find new difficulties. An overview of the whole process can be seen in figure \ref{fig:KundischProcess}.

The ETDP begins with the clear identification and motivation of a problem (I), which involves specifying the observed phenomenon (Step 1), defining the taxonomy’s target users (Step 2), and articulating the taxonomy’s purpose(s) (Step 3). These purposes commonly fall into two categories: purely structuring or combined identification and structuring. Importantly, an entry or re-entry point exists before Step 2 to allow for refinement of the target user scope during subsequent demonstration or evaluation phases, reflecting whether the taxonomy is suitable for part or all of the originally defined audience or beyond. 

The second step of the design process is to define the objectives of the solution (II). To actively steer the design of the taxonomy Nickerson et al.(2022) introduce the so called meta-characteristic which serves as "a basis for the choice of characteristics in the taxonomy". The meta-characteristic in its turn is based on the intended purpose of the taxonomy as defined in step one. The iterative nature of the whole process necessitates ending conditions as a basis for the decision when to stop the process. \cite{nickerson_method_2013} provides a list of ending conditions which they categorize into subjective ones (concerning the usefulness of the taxonomy) and objective ones (concerning the validity of the taxonomy).
The ETDP emphasizes determining the meta-characteristic and both objective and subjective ending conditions upfront, alongside establishing evaluation goals to align the problem space with the solution space effectively. These evaluation goals typically focus on enhancing the description, identification, classification, analysis, or clustering of objects representing the phenomenon in comparison to approaches lacking a taxonomy. The design of a taxonomy is heavily influenced by the point of view chosen by the researchers. Consequently, taxonomies should not be judged if they are 'right' or 'wrong' but if they are suitable for their intended purpose. 

The third step consists of the actual development of the taxonomy (III). Throughout this step, taxonomy operations — such as adding, updating, or deleting characteristics and dimensions — facilitate the structural evolution of the taxonomy as needed.
These operations can be based on either an empirical-to-conceptual or a conceptual-to-empirical approach, chosen based on the domain knowledge and data availability. When data are limited but domain understanding is strong, a conceptual-to-empirical approach is recommended, whereas extensive data with limited domain insight favors the empirical-to-conceptual approach. In cases of balanced knowledge and data, researchers must use judgment to select the method, while insufficient knowledge and data necessitate further domain exploration. The empirical-to-conceptual method involves selecting a representative object subset, identifying discriminatory characteristics derived from a meta-characteristic, and grouping these into mutually exclusive, collectively exhaustive dimensions to form an initial taxonomy. Conversely, the conceptual-to-empirical approach starts with hypothesizing dimensions based on expert judgment, which are then validated against actual objects. These two approaches are not mutually exclusive and it can be helpful to switch between the two between iterations.

The process culminates in demonstration and rigorous analysis, guided by objective ending conditions (IV) that verify whether the taxonomy fulfills essential criteria and subjective conditions ensuring its practical usefulness. Evaluation activities (V) include assessing the taxonomy’s constructs and their relationships, and are often conducted via consensus methods among researchers and stakeholders. The importance of thorough ex-post evaluation as introduced by \cite{kundisch_update_2022} is underscored by the statement in \cite{nickerson_method_2013} that they "are not able [..] to give sufficient conditions [for a taxonomy to be useful] other than to say that a taxonomy is useful if others use it". It can be achieved by challenging and validating the artefact’s utility in supporting target users’ achievement of intended goals. Finally, clear communication (VI) of the taxonomy development process and its final product is advocated, incorporating appropriate visualizations tailored to the target users and purposes, as well as detailed descriptions of each characteristic and dimension. Collectively, ETDP’s iterative, flexible, and user-centered methodology provides comprehensive operational support, enabling the creation of robust and valuable taxonomies that can evolve over time in response to new insights and application contexts.

\section{Results} 
\subsection{Problem Specification} 
There is a big body of research on difficulties, misconceptions and errors of novice programmers. There have also been some attempts at creating collections and inventories of difficulties \cite{chiodini_curated_2021, sorva_visual_2012, cortinovis_open_2023}, but beyond sorting them to bigger concepts (like selection, iteration or object-oriented programming) most of them have not attempted to create a finer grained classification. \cite{chiodini_curated_2021} have collected a lot of information about some difficulties like their possible origin and symptoms but have not used this information to classify these difficulties. With the taxonomy presented here we build on this prior work and try to fill the gap of a fine-grained classification for a specific kind of problems, namely the ones concerning control flow and the basic control structures of selection and iteration.

\subsection{Target User Group}
The taxonomy presented here is mainly aimed at other computing education researchers that are investigating difficulties - especially concerning reading code and control flow but not exclusively. This taxonomy can help integrate new results into the existing body of research on the topic, such as checking if a discovered error is actually a novel finding or if it is just a new facet of an already known difficulty. It is also intended to serve as basis for a theory that can explain in a concise way how students learn about control flow and where and why they struggle. It can even give ideas to researchers and educators for how to improve teaching.

\subsection{Intended Purpose}
The purpose of this taxonomy is to harmonize the results on understanding code with simple control structures from different sources, to serve as a tool to categorize students' errors effectively and unambiguously, and to serve as a basis for a deeper understanding of student difficulties. It therefore has a mostly structuring purpose, though the emerging structure can potentially give hints to where to look for undiscovered difficulties, help explain the origins of difficulties and how to overcome them.

\subsection{Meta Characteristic} \label{metacharacteristic}
As our goal for this taxonomy is to help understand the difficulties students have and the relationship between these difficulties, we chose the \textit{characteristics of the computing concepts that students fail to understand correctly} as the meta characteristic, since this is what ideally students are supposed to understand after an introductory programming course. When we can classify which underlying concept students struggle with, we might be able to better address these issues.

\subsection{Ending Conditions} \label{endingconditions}
We chose the following objective ending conditions from the examples given by \cite{nickerson_method_2013}:

\begin{enumerate}
    \item All objects or a representative sample of objects have been examined
    \item No object was merged with a similar object or split into multiple objects in the last iteration
    \item No new dimensions or characteristics were added in the last iteration
    \item No dimensions or characteristics were merged or split in the last iteration
    \item Every characteristic is unique within its dimension
    \item Each cell (combination of characteristics) is unique and is not repeated
\end{enumerate}

As subjective ending conditions we have chosen the ones proposed by \cite{nickerson_method_2013}(our additions are in square brackets):
\begin{enumerate}
    \item Concise: Does the number of dimensions [and characteristics per dimension] allow the taxonomy to be meaningful without being unwieldy or overwhelming?
    \item Robust: Do the dimensions and characteristics provide for differentiation among objects sufficient to be of interest? 
    \item Comprehensive: Can all objects within the domain of interest be classified? Are all dimensions of the objects of interest identified?
    \item Extendable: Can a new dimension or a new characteristic of an existing dimension be easily added?
    \item Explanatory: [...] Do the dimensions and characteristics explain [enough] about an object?
\end{enumerate}

\subsection{Evaluation goals} \label{evaluationgoals}
An evaluation is a field test. So to evaluate if our taxonomy serves its intended purpose we will give the taxonomy to researchers not involved in the design process to look at student answers to exam questions where students need to read small programs and predict their output, and see how well these researchers can categorize these answers into the created taxonomy. Feedback from these researchers on the comprehensiveness, ease of use and understandability of the taxonomy will serve as basis for further design iterations.

\subsection{Building Approach} 
According to \cite{kundisch_update_2022} there is an empirical-to-conceptual approach and a conceptual-to-empirical approach to building a taxonomy. Both should be incorporated into the design process. Depending on the purpose of the taxonomy (descriptive vs. normative) and the existence of a relevant knowledge base one should start with one or the other. The purpose of our taxonomy is on one hand descriptive as it describes something (student difficulties) as it is and tries to give it a framework that helps analyzing it. On the other hand, it has a normative aspect as our approach emphasizes the desired final state (correct understanding of the concept of control flow and the necessary subconcepts) and is intended to help achieve that state. So, the purpose does not give a conclusive decision on which approach to choose. Therefore, we look at the existing knowledge base for student errors. As there is already a substantial body of research done on student difficulties, we opt for a conceptual-to-empirical approach to start with as recommended by \cite{nickerson_method_2013}. 

We diverge from the method of \cite{nickerson_method_2013} and \cite{kundisch_update_2022} in that our taxonomy does not consist of orthogonal dimensions that are independent of each other but of a hierarchy of dimensions that divide the space of errors into ever finer-grained categories. The reason for this is, that we - as opposed to the aforementioned authors - do not follow a phenetic approach. In phenetics, classification is based on the similarities of the objects to be classified. In our case this would mean looking at the similarities of student answers and cluster them with the help of statistical techniques to quantify the similarities. But as we know that on one hand many different difficulties can lead to similar or even the same student answers and on the other hand such a classification would not be very helpful for understanding students' thought processes we choose a different approach. Alternatively, cladistic classification in biology is based on the evolutionary relationship among organisms and results in the well known evolutionary trees. In our context this would mean categorizing according to the origins of students' errors, e.g. misconceptions from natural language, everyday life or math education. This could be an extension of our taxonomy or even a new taxonomy of its own. Our point of view as expressed in our meta characteristic in section \ref{metacharacteristic} is defined by the underlying concepts students need to understand in order to master the subject. This is quite close to cladistics, which is why we chose our structure to be similar to an evolutionary tree. Apart from this we still stuck with the process proposed by \cite{kundisch_update_2022} as the rest of the design process was still suitable.

\section{Taxonomy} 

The first obvious choice for a dimension is the control structure the difficulty is concerned with. So far, the possible characteristics in our taxonomy are selection and iteration. This dimension is easily extensible to e.g. sequence or subroutine. On the next lower level, we categorized the difficulties into groups that concerned the same aspect of the control structure (denoted with arabic numerals below) - or in other words \textit{what} they do not understand. On the third level and beyond (denoted with lowercase latin letters, lowercase roman numerals, uppercase latin letters and uppercase roman numerals) we discriminated according to \textit{how} the students have difficulties understanding the concept. This can be a clear misconception, fragile knowledge or just a description of how students struggle with the concept. 

This categorization proved to be more difficult than expected. The description of errors in existing research show a wide variety of detail, granularity and aspects under consideration. Therefore, we had to first decide which descriptions are on a higher level of classification and which ones are on a lower one. Additionally, we had to compare descriptions of difficulties that are similar to each other and decide wether they actually are distinct from each other or are just different descriptions of the same phenomenon. Finally, certain descriptions were so broad and superficial that a further breakdown into new subcategories was needed. 

The basis for our design were on one hand the errors, difficulties and misconceptions that we know from our own research where we conducted semi-structured interviews and written tests with students in various introductory programming courses. On the other hand, we drew from existing literature - though we do not claim to have incorporated every paper on difficulties about control structures yet. 

After several iterations, we came up with the structure presented in Sections \ref{Selection} and \ref{Iteration}. Where the categories correspond with descriptions found in literature, we added the reference and in some cases the original wording. Thus the categories without references are newly created categories or even newly found difficulties from our own research.

\subsection{Selection} \label{Selection}
This section describes the part of the taxonomy about errors concerning selection statements.

\begin{enumerate}
\item Selective nature of selection statements:

The errors in this category are concerned with the very core concept of conditional statements. They occur when students do not (fully) understand that in a conditional statement a program "decides" if a certain part of the code is executed or not or they do not understand how this happens.
    \begin{enumerate}
    	\item Not considering the truth of a control condition as the cause for the execution of its branch
        \begin{enumerate}
            \item Thinking that all branches are executed independent of their condition. \cite{sleeman_pascal_1986}
            \item Thinking that branches are executed due to something other than the truth of the condition (e.g. the "intention" of the programmer or part of the code outside of the selection statement). \cite{sychev_explain_2023} proposes two separated errors, but we consider the underlying difficulty to be essentially the same: "Executing a wrong branch after checking a condition: The student thinks that a different branch can be executed after a selection-statement condition evaluates to true." and "Starting a \textit{then}-branch after a false condition: The student tries to start a selection-statement branch even though its condition is false."
        \end{enumerate}
    	\item Misunderstanding exit of selection statement
        \begin{enumerate}
            \item Thinking that a selection statement is not exited after one branch of a chaining conditional ("else-if") has been executed. \cite{sychev_explain_2023}: "Two or more branches of one selection statement are executed if they are true: The student thinks that more than one branch can be executed during one execution of a selection statement."
            \item Thinking that a selection statement is exited BEFORE a branch has been executed or end of statement has been reached: Not checking all conditions until one is correct or end of selection statement reached. \cite{sychev_explain_2023}: "Exiting a selection without executing a branch or checking all the conditions: The student thinks that a selection statement can finish before checking all its conditions."
        \end{enumerate}
        \item Expecting the body of a selection statement to be executed if condition is wrong
    \end{enumerate}
    
\item Boundaries of body of selection statement:

Here students do not know what parts of a program are part of a selection statement and which ones are not.
	\begin{enumerate}
        \item Thinking that a part of the code which is placed after the selection body belongs to its branch. \cite{sychev_explain_2023}: "Inside a selection-statement branch, the student executes an action that is placed after the selection statement."
        \item Thinking that a part of the code which is placed inside the selection body is belongs to it.
	\end{enumerate}
    
\item \textit{Else}-branch:

This category comprises errors that are concerned with the \textit{else}-branch.
	\begin{enumerate}
	    \item Not understanding the necessity of the \textit{else}-branch
        \begin{enumerate}
            \item Thinking that all selection statements require two cases, one to execute if the condition is true, the other (the \textit{else}-branch) to execute if the condition is false. \cite{chiodini_curated_2021} 
        \end{enumerate}
        \item Expecting undefined or extreme behavior in case of evaluation of condition to false \cite{chiodini_curated_2021}
        \begin{enumerate}
            \item Thinking that control goes back to the beginning of the program when the condition is false. \cite{putnam_summary_1986, swidan_programming_2018, sorva_visual_2012}
            \item Thinking that a false condition ends the program if there is no \textit{else}-branch. \cite{putnam_summary_1986, sleeman_pascal_1986, swidan_programming_2018, sorva_visual_2012}
        \end{enumerate}
        \item Confusion between "if(C) A; else B;" and "if (C) A; B;"
        \begin{enumerate}
            \item "if(C) A; else B;" is interpreted as  "if(C) A; B;". In other words, the \textit{else}-branch always executes independent of the value of the condition. \cite{chiodini_curated_2021}
            \item "if(C) A; B;" is interpreted as "if(C) A; else B;". In other words, using \textit{else} is optional or the next statement is always the \textit{else}-branch. \cite{sleeman_pascal_1986, du_boulay_difficulties_1986, sorva_visual_2012}
        \end{enumerate}
	\end{enumerate}
    
\item Timing of Condition Check:

This category lists all errors where students do no know that there is a precise timing when the condition is checked or when this happens. 
	\begin{enumerate}
	    \item Thinking an \textit{if}-statement or parts of it are executed multiple times
        \begin{enumerate}
            \item Multiple checks of the condition: The student thinks that in a selection statement, a branch condition can be evaluated several times. \cite{sychev_explain_2023}
            \item Multiple execution of whole \textit{if}-statement: The student thinks that a selection statement can have several iterations during a single execution of the statement...
            \begin{enumerate}
                \item …until the condition is wrong: The body of the \textit{if}-statement executes as many times as needed to ensure the condition does not hold anymore and execution can continue. \cite{chiodini_curated_2021, bastian_misconceptions_2025}
                \item …for number of times \cite{sychev_explain_2023} determined by something else than the truth of the condition e.g. by purpose of program. 
            \end{enumerate}
        \end{enumerate}
        \item Thinking the condition is constantly being checked parallel to the execution of the program. In this misconception an \textit{if}-statement gets executed as soon as its condition becomes true. \cite{pea_language-independent_1986} Different description by \cite{cortinovis_open_2023}: "Confusion between IF <COND> and ON/WHEN <COND> (parallelism bug)" 
        \begin{enumerate}
            \item Thinking that the condition is only checked in the part of the program after the selection statement.
            \item Thinking that the condition is checked in the whole program.
        \end{enumerate}      
        \item Thinking the condition is checked in hindsight
        \begin{enumerate}
            \item Thinking that a condition can be evaluated after the corresponding branch of the selection statement is completed and that if the condition is true AFTER executing the corresponding branch, the branch is executed. \cite{sychev_explain_2023}
            \item If condition becomes true from a certain point INSIDE of the selection-body…
            \begin{enumerate}
                \item …the body is executed from this point on forward (part of selection-body before this point is not executed) either…
                
                     (I) …to the end of the branch.
                     
                     (II) …as long as the condition remains true. As soon as the condition does not hold anymore the conditional statement is left.
                \item …the whole body is executed.
            \end{enumerate}
        \end{enumerate}
    	\item Misunderstanding timing of condition check in multi-branch selection statement
        \begin{enumerate}
            \item Thinking that the conditions of a multi-branch selection statement are checked in the wrong order: The student thinks that, in a multi-branch selection statement, conditions can be checked not in the order they are placed in the statement \cite{sychev_explain_2023, albrecht_sometimes_2020}
            \item Confusion between "if (c) A; else B;" and "if (c) A; if (!c) B;". \cite{chiodini_curated_2021} interpret this as one single misconception. We deem it helpful to separate it into two distinct categories:
            \begin{enumerate}
                \item The student interprets "if(C) A; else B;" as "if(C) A; if(!C) B;". Condition is checked again for the \textit{else}-branch after execution of the \textit{then}-branch and if the condition is wrong the \textit{else}-branch is executed. 
                \item The student interprets "if(C) A; if(!C) B;" as "if(C) A; else B;". Condition of sequential selection statements with opposite conditions are not checked separately but the second one is treated as an \textit{else}-branch. Therefore, change of value of condition through either statements in the body of selection statement or through the evaluation of the condition is ignored.
            \end{enumerate} 
        \end{enumerate}
	\end{enumerate}
    
\item Nesting:

When a conditional statement occurs within another control structure we speak of nesting. This nested constructs tend to be difficult for students to understand.
    \begin{enumerate}
        \item "Confusion between sequencing versus nesting of \textit{if}-statements." \cite{cortinovis_open_2023}
        \begin{enumerate}
            \item Misunderstanding nesting as sequencing. Thinking that when two \textit{if}-statements are nested and the condition of the outer if-statement is wrong but the condition of the inner \textit{if}-statement is correct, the body of the inner if-statement is still executed
            \item Sequencing is nesting. When a program contains two \textit{if}-statements right next to each other both conditions need to be true for the body of the second \textit{if}-statement to execute.
        \end{enumerate}
        \item Ignoring \textit{if}-condition inside of loop. The student thinks that the condition of an \textit{if}-statement inside of a loop does not need to be evaluated and the body executes in any case.
    \end{enumerate}

\end{enumerate}

\subsection{Iteration} \label{Iteration}
This section describes the part of the taxonomy about errors concerning iteration statements.


\begin{enumerate}

    \item Repeating nature of iteration statements:

    The errors in this category are concerned with the very core concept of iteration statements. They occur when students do not (fully) understand that in an iteration statement a part of the program is repeated a certain number of times. This number can vary form 0 to infinity (theoretically). Errors where students do not understand this or do not understand how a program "decides" upon the number of repetitions belong into this category.
    \begin{enumerate}
        \item Belief that the loop body is always executed once, independent of the condition being true or not. \cite{grover_measuring_2017,eckert_student_2022} This is the same as completely ignoring the loop construct except for the body.
        \item Believing the body of the \textit{for}-statement executes either one or zero times depending on whether the condition holds or not. Another description of this is "loop is if". \cite{chiodini_curated_2021, muhling_design_2015, bastian_misconceptions_2025, eckert_student_2022}
        \item Initialization, condition and update of control variable are understood in a mathematical sense as equations that need to be fulfilled or are "active" at the same time. \cite{eckert_student_2022}
        \item Thinking the counter of a loop with a fixed number of repetitions represents how many of the statements inside the loop body are executed. \cite{bastian_misconceptions_2025}
    \end{enumerate}
    \item Interaction with code outside of iteration statement:

    This category comprises errors concerning the interaction between iteration statements and the rest of the code.
	\begin{enumerate}
        \item Misunderstanding boundaries of body of iteration statement
        \begin{enumerate}
            \item Part of body of iteration statement is considered as not belonging to it \cite{sleeman_pascal_1986}
            \item Part of code after iteration statement is considered to be part of body of iteration statement \cite{putnam_summary_1986, sleeman_pascal_1986, eckert_student_2022, swidan_programming_2018, sychev_explain_2023, bastian_diagnose_2021}
        \end{enumerate}
        \item Misunderstanding behavior of Code outside of loop
        \begin{enumerate}
            \item Believing that the program terminates after loop finishes. Consequently the code after a loop is never executed. \cite{du_boulay_difficulties_1986, bastian_diagnose_2021}
            \item Understanding a \textit{for}-loop as some sort of iterative \textit{if-else}-construct.  If the condition evaluates to 'true' the loop body is iterated, otherwise - if it evaluates to 'false' - the adjacent code is iterated.
        \end{enumerate}
    \end{enumerate}
    \item Order of execution of loop parts:

    A loop consists of 4 parts: Initialization, condition check, body and update of control variable. The order in which they are executed is clearly defined. The errors that occur when students have difficulties with this order are grouped into this category.
    \begin{enumerate}
        \item Skipping a part
        \begin{enumerate}
            \item Skipping the execution of the loop body even though the loop condition is true. The student thinks that the computer can proceed to the next action of the loop (e.g., checking the condition again) without executing the loop’s iteration. \cite{sychev_explain_2023}
            \item Directly executing the loop body after initializing the control variable. The first iteration is executed independently of the boolean value of the condition. This is a special case, as this is the correct conception of a \textit{do-while}-loop, but a misconception for \textit{while}- and \textit{for}-loops. \cite{sychev_explain_2023}
            \item Skipping the condition check of a loop. The student thinks that the condition check of a loop can be omitted. \cite{sychev_explain_2023}
            \item Skipping the update step of a loop. The student thinks that the update step of a loop can be omitted after an iteration finishes. \cite{sychev_explain_2023}
            \item Skipping the initialization of a loop: The student thinks that the initialization step of a loop should not be executed when the loop starts. \cite{sychev_explain_2023}
        \end{enumerate}
        \item Misconceptions about the order of loop-parts
        \begin{enumerate}
            \item Thinking that a loop is initialized after its iteration. The student thinks that the initialization step of a loop can be executed after the iteration finishes. \cite{sychev_explain_2023}
            \item Thinking that the update of the control variable happens before the loop body executes
            \begin{enumerate}
                \item with any control variable update.
                \item only with "++" pre-increment. Thinking that using a "++" pre-increment operator in a loop’s update part, (e.g. "for(...; ...; ++i)" ...) means that the increment happens before the loop body executes, while using a "++" post-increment (e.g. "for(...; ...; i++)" ...) means that the increment happens after the loop body executes.
            \end{enumerate}
            \item Thinking that the update of the control variable happens before the loop body AND before the condition check i.e. the first thing that happens.
            \item Thinking that the condition is permanently checked also during the execution of the loop body and the iteration terminates as soon as condition changes to false. \cite{pea_language-independent_1986, du_boulay_difficulties_1986, swidan_programming_2018, bastian_misconceptions_2025}
            \item Grouping of the statements inside the loop body: When there are for example two statements in the loop body, the student assumes that the first statement is executed as often as specified by the loop and only then the second is executed as often as specified. \cite{grover_measuring_2017}
        \end{enumerate}
    \end{enumerate}
    \item State:
    
    Even though that we limit ourselves to the concept of control flow in this paper, there are difficulties with the state of the program, that cause problems with the control flow.
    \begin{enumerate}
        \item Difficulties concerning the control variable
        \begin{enumerate}
            \item Students have difficulties in understanding changes to loop control variables. Another description of this is "Improper handling of loop counter". \cite{du_boulay_difficulties_1986, cortinovis_open_2023, sorva_visual_2012} A more general description by \cite{cortinovis_open_2023}: "Incorrect update of condition in conditional loops."
            \begin{enumerate}
                \item Inconsistent change of update "direction": The student thinks that during the execution of the loop the update to the control variable changes to from iteration to iteration even though it is constant. \cite{eckert_student_2022}
                \item Thinking that an iteration statement has a default behavior of just incrementing or decrementing by 1, regardless of how the update actually looks like. 
                \item Increment "++" misinterpreted as incrementing by 2.
            \end{enumerate}
            \item Mental disconnection of control variable in loop body, in loop condition and in update statement.
            \begin{enumerate}
                \item Thinking manipulation of control variable inside loop body has no effect on evaluation of condition. \textit{for}-loop control variables do not have values inside the loop or their values can be arbitrarily changed. \cite{putnam_summary_1986, cortinovis_open_2023, sorva_visual_2012}
                \item Thinking that the update of the control variable has no influence on the behavior of the body. This misconception can express itself by student expecting that a loop produces the exact same output for each iteration and not accounting for changing variables that can also change the output for different iterations \cite{sleeman_pascal_1986, sorva_visual_2012}
            \end{enumerate}
            \item Thinking that a loop control variable constrains the values handled within the loop. \cite{sleeman_pascal_1986, putnam_summary_1986, sorva_visual_2012} 
            \begin{enumerate}
                \item The control variable constraints the values that can be read via input. 
                \item The control variable constraints the values that can be printed.
            \end{enumerate}
        \end{enumerate}
        \item Difficulties concerning other parts of the state: Changing the state (excluding control variable) does not affect behavior of body. Other variables and data structures can change their value during the execution of an iteration of a loop. Some students do not understand that can affect the execution of coming iterations.
    \end{enumerate}
	
    \item Difficulties with loop condition:
    
    This category comprises all errors where students have difficulties about the effect of the condition and how it is evaluated.
	\begin{enumerate}
	    \item Misunderstanding effect of the condition
        \begin{enumerate}
            \item Thinking that the number of iterations is determined by something else than the loop-condition (e.g. the "intention" of the programmer or part of the code outside of the selection statement). Therefore a loop can also be continued after the condition evaluates to false. \cite{sychev_explain_2023}
            \item The condition of an iteration statement describes the terminating case, thus one exits the loop when it becomes true and enters when it is false... 
            \begin{enumerate}
                \item …only if the condition is wrong to begin with. However if it is true when the condition is first checked, one exits the loop when it becomes false. 
                \item …in any case. \cite{sychev_explain_2023, chiodini_curated_2021}
            \end{enumerate}
            \item Misconception that when the condition is wrong to begin with, but becomes true after a number of iterations the condition is "validated" in hindsight. 
            \begin{enumerate}
                \item Student believes that the loop is iterated until the condition is true and then is executed until condition is wrong \textit{again}.
                \item Student believes that the iterations until the condition is true are skipped and then the loop is executed until condition is wrong \textit{again}.
            \end{enumerate}
        \end{enumerate}
        \item Misunderstanding the evaluation of the condition
        \begin{enumerate}
            \item Struggle to identify the correct range of the control variables of loops, based on understanding of operators. This category could be considered as a lexical difficulty, we kept it in the taxonomy anyway, as it also encompasses the conceptual understanding of condition evaluation.
            \begin{enumerate}
                \item Students confuse the meaning of the strict and non-strict relational operators ("<" and "<=" or ">" and ">="). \cite{eckert_student_2022, grover_measuring_2017, albrecht_sometimes_2020}
                \item Students confuse operands "<" and ">".
            \end{enumerate}
            \item Students make logical errors while evaluating the condition \cite{radakovic_common_2024}
        \end{enumerate}
	\end{enumerate}
    \item Nesting

    As described above nesting is difficult concept for students, even more so for iteration than for selection.
    \begin{enumerate}
        \item Belief that nested loops are executed simultaneously as if they were one loop with two control variables and two termination conditions that are changed and checked simultaneously. \cite{cetin_students_2015, eckert_student_2022}
        \item Misconception that two nested loops act like two loops that are executed sequentially. other description: Confusion between sequencing versus nesting of iterative statements. \cite{cetin_students_2015, mladenovic_comparing_2018, cortinovis_open_2023, bastian_diagnose_2021}
        \item Switching inner and outer loop. The students think that the outer loop is executed completely for every iteration of the inner loop.
        \item Ignoring the outer loop. The student thinks that no repetitions of the outer loop takes place. \cite{bastian_misconceptions_2025}
        \item Thinking the counter of the inner loop with a fixed number of repetitions overwrites the outer counter and the statements are executed as one loop body. \cite{bastian_misconceptions_2025}
        \item Believing part of the outer loop is not executed after the inner loop, if inner loop is not executed at all. The Student thinks that the next iteration of the outer loop directly starts, if the condition of the inner loop is wrong in its first condition check. \cite{eckert_student_2022}
    \end{enumerate}

    \item Break and continue:

    Even though they are not taught in every introductory programming class and are sometimes even deemed bad code style, \textit{break} and \textit{continue} remain present in many programming languages and students tend to struggle with their meaning.
    \begin{enumerate}
        \item Misunderstanding the \textit{break}-statement
        \begin{enumerate}
            \item Student thinks the \textit{break} statement means that the next iteration of the loop is started. ("Break is continue")
            \item Thinkin that \textit{break} terminates the whole program.
            \item Student thinks \textit{break} means to continue to execute the loop body. This is equivalent to ignoring the \textit{break} statement.
        \end{enumerate}
        \item Misunderstanding the \textit{continue}-statement
        \begin{enumerate}
            \item Student thinks \textit{continue} means to continue to execute the loop body. This is equivalent to ignoring the \textit{continue} statement. 
        \end{enumerate}
    \end{enumerate}
\end{enumerate}

\section{Discussion}
In this section we want to discuss if our taxonomy satisfies our ending conditions as described in section \ref{endingconditions}. We also describe how the taxonomy can be used to clarify complex errors. Finally we discuss the limitations of the taxonomy.

\subsection{Check of objective ending conditions}

We see objective ending condition (1) fulfilled even though there are certainly still difficulties in literature that we have not considered yet, we have incorporated every difficulty that we found in a subset of important papers as well as our own research as long as they lay within our scope as described in section \ref{scope}. The ending conditions (2)-(4) have been achieved when after several iterations we did not see the need for further changes to the taxonomy in the last iteration. Conditions (5) and (6) are self-evidently true as can be seen from the taxonomy itself.

\subsection{Check of subjective ending conditions}

We think that the taxonomy allows for a meaningful distinction while giving a clearer picture and thereby reducing the overwhelming variety of errors. Therefore, we think the taxonomy fulfills the subjective ending condition (1). The differentiation between the errors based on the conceptual aspects and how students struggle with them should be of great interest for researchers and teachers alike, as it can help paint a clearer picture where the bigger problems lie (ending condition (2) and (5)). Ending condition (3) is satisfied as far as errors know to us are concerned. The ability of the taxonomy to accommodate for further types (ending condition (4)) of errors will be put to the test in further evaluation steps.

\subsection{Usage example}

\begin{figure}[!ht]
    \centerline{\includegraphics[width=4in]{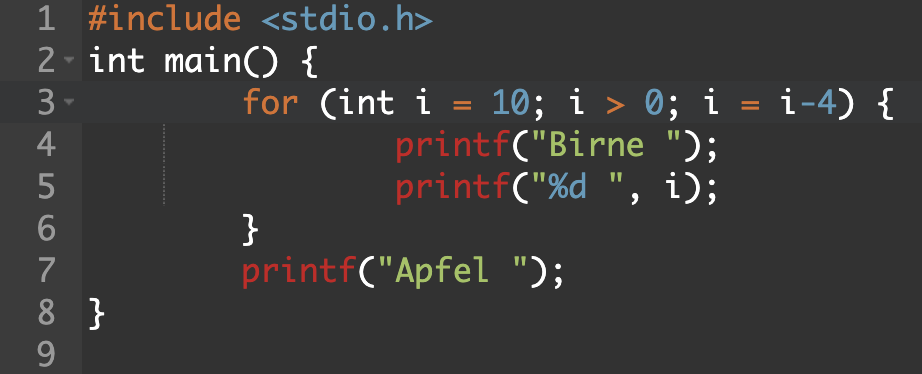}}
    \caption{Students were given this piece of code and asked to predict the output.}
    \label{fig:taskexample}   
\end{figure}

\begin{figure}[htbp]
  \centering
  \begin{subfigure}[b]{0.2\textwidth}
    \includegraphics[width=\linewidth]{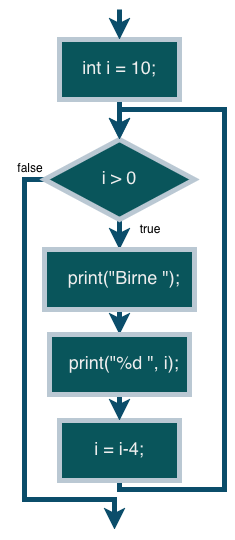}
    \caption{Correct conception}
    \label{fig:fig1}
  \end{subfigure}
  \hfill
  \begin{subfigure}[b]{0.2\textwidth}
    \includegraphics[width=\linewidth]{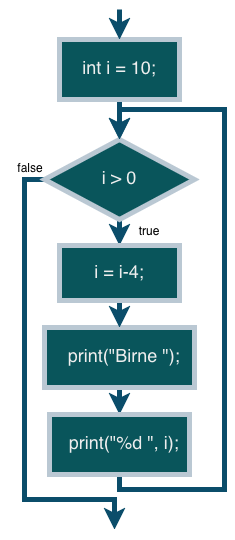}
    \caption{Difficulty Iteration-(3)-(b)-(ii)-(A)}
    \label{fig:fig2}
  \end{subfigure}
  \hfill
  \begin{subfigure}[b]{0.2\textwidth}
    \includegraphics[width=\linewidth]{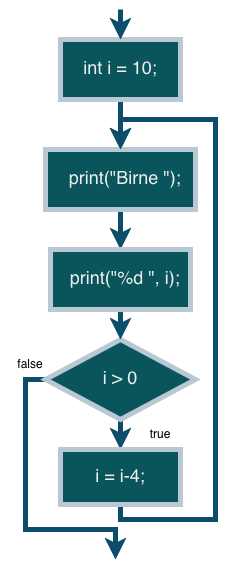}
    \caption{Difficulty Iteration-(3)-(b)-(ii)-(A) AND Iteration-(3)-(a)-(ii)}
    \label{fig:fig3}
  \end{subfigure}
  \hfill
  \begin{subfigure}[b]{0.2\textwidth}
    \includegraphics[width=\linewidth]{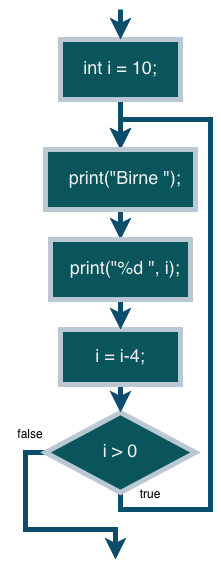}
    \caption{Difficulty Iteration-(3)-(a)-(ii)}
    \label{fig:fig4}
  \end{subfigure}
  \caption{Different (mis)conceptions of students about a simple loop.}
  \label{fig:conceptions}
\end{figure}

In order to illustrate the use of the taxonomy we analyze a common error that we found multiple times in our own research that seemed difficult to decode at first. The students were given a piece of code (see figure \ref{fig:taskexample}) and asked to predict the output. Some students gave the correct answer (\textit{"Birne 10 Birne 6 Birne 2 Apfel"}). Many others gave one of the two following wrong answers: \textit{"Birne 6 Birne 2 Birne -2 Apfel"} or \textit{"Birne 10 Birne 6 Birne 2 Birne -2 Apfel"}. The students with the correct answer probably have the correct mental model of a loop shown in figure \ref{fig:fig1}. Students with the misconception Iteration-(3)-(b)-(ii)-(A) shown in figure \ref{fig:fig2} consequently choose the answer \textit{"Birne 6 Birne 2 Birne -2 Apfel"} as they believe the update step happens before the execution of the body. Now for the answer \textit{"Birne 10 Birne 6 Birne 2 Birne -2 Apfel"} it is a bit mor complicated as students choosing this answer have two separate misconceptions that overlap. On one hand they also have the misconception Iteration-(3)-(b)-(ii)-(A) but at the same time also have the misconception Iteration-(3)-(a)-(ii) resulting in the mental model shown in figure \ref{fig:fig3}. Students that only have the misconception Iteration-(3)-(a)-(ii) have a mental model as shown in figure \ref{fig:fig4} which in this case leads to the correct answer (the misconception is not visible in this case). What we like to show with this example is that the clear definition of difficulties helps us categorize errors more clearly. Here four seemingly different errors actually consist of three mental models and one combination of two of them.

\subsection{Limitations}
As mentioned in section \ref{evaluationgoals} the taxonomy needs to be tested on students' errors that are not part of the error samples used to generate the taxonomy in the first place. This is necessary to make sure that the taxonomy is useful in categorizing students' errors and that it is not overfitted to the sample of errors that we looked at. 
Another necessary step towards a completely evaluated taxonomy is to test if the categories are clearly intelligible by researchers and teachers. This will be done by inviting researchers and teachers to apply the taxonomy in their own contexts and report back on their experience.
Both these steps are planned for future research.

\section{Conclusion}
This paper has presented a comprehensive taxonomy that systematically categorizes the difficulties novice programmers encounter when reading and trying to understand the control flow constructs 'selection' and 'iteration'. By leveraging the Extended Taxonomy Design Process (ETDP), we have developed a flexible framework that integrates both empirical evidence and conceptual insights from existing literature and original research. Our classification reveals the complexity underlying common programming errors and offers a valuable tool for computing education researchers and educators to better identify, describe, and address these challenges. While the taxonomy satisfies both objective and subjective ending conditions for robustness, comprehensiveness, and extensibility, further empirical evaluation is needed to confirm its practical utility and understandability in diverse educational contexts. Future research will focus on validating the taxonomy with new student data and gathering feedback from the research and teaching community to refine and extend its applicability. Ultimately, this taxonomy aims to contribute to improved instructional design and support more effective learning of foundational programming concepts related to control flow.


\bibliographystyle{ACM-Reference-Format}
\bibliography{ICER_2026}


\end{document}